\documentclass[conference]{IEEEtran}
\IEEEoverridecommandlockouts
\usepackage{cite}
\usepackage{amsmath,amssymb,amsfonts,amsthm}
\usepackage{graphicx}
\usepackage{textcomp}
\usepackage{xcolor}

\usepackage{bm}
\usepackage{mathtools}
\usepackage{float}
\usepackage[ruled,vlined,linesnumbered]{algorithm2e}
\usepackage{algpseudocode}
\usepackage{leftidx}
\usepackage{url}
\usepackage{tabularx, multirow, multicol}
\usepackage{color}
\usepackage{soul}
\usepackage{caption}
\usepackage{subcaption}
\usepackage{float}
\usepackage{balance}
\usepackage{setspace}
\usepackage{bbm}

\usepackage{pifont}
\newcommand{\cmark}{\ding{51}}%
\newcommand{\xmark}{\ding{55}}%

\theoremstyle{definition}
\newtheorem{definition}{Definition}

\newcolumntype{B}{>{\centering\arraybackslash}m{0.27\linewidth}}
\newcolumntype{L}{>{\raggedright\arraybackslash}m{0.27\linewidth}}
\newcolumntype{X}{>{\centering\arraybackslash}m{0.1\linewidth}}
\newcolumntype{R}{>{\raggedleft\arraybackslash}m{0.095\linewidth}}
\newcolumntype{S}{>{\centering\arraybackslash}m{0.085\linewidth}}
\newcolumntype{T}{>{\centering\arraybackslash}m{0.135\linewidth}}

\newenvironment{talign*}
 {\let\displaystyle\textstyle\csname align*\endcsname}
 {\endalign}

\newcommand{\autophrase}{AutoPhrase\xspace}

\newcommand{\lof}{LOF\xspace}
\newcommand{\acs}{ANCS\xspace}
\newcommand{\rshash}{RS-Hash\xspace}

\newcommand{\tonmf}{TONMF\xspace}

\newcommand{\cvdd}{CVDD\xspace}
\newcommand{\cvddd}{CVDD\textsubscript{d}\xspace}
\newcommand{\cvddw}{CVDD\textsubscript{w}\xspace}

\newcommand{\vmfd}{vMF\textsubscript{d}\xspace}
\newcommand{\vmfw}{vMF\textsubscript{w}\xspace}
\newcommand{\kjnn}{kj-NN\xspace}
\newcommand{\westclass}{WeST-Class\xspace}
\newcommand{\smclass}{SM-Class\xspace}

\newcommand{\proposed}{OOCD\xspace}
\newcommand{\proposedd}{OOCD\textsubscript{d}\xspace}
\newcommand{\proposedw}{OOCD\textsubscript{w}\xspace}

\newcommand{\nyt}{NYT\xspace}
\newcommand{\arxiv}{arXiv\xspace}

\DeclareMathOperator*{\argmax}{argmax}

\newcommand{\wvec}[1]{\bm{w}_{#1}}
\newcommand{\vvec}[1]{\widetilde{\bm{w}}_{#1}}
\newcommand{\dvec}[1]{\bm{d}_{#1}}
\newcommand{\cvec}[1]{\bm{c}_{#1}}

\newcommand{\ptloss}{\mathcal{L}_{pretrain}}
\newcommand{\stloss}{\mathcal{L}_{refine}}

\newcommand{\wordset}{\mathcal{W}}
\newcommand{\docuset}{\mathcal{D}}
\newcommand{\cateset}{\mathcal{C}}

\newcommand{\smallsection}[1]{{\vspace{0.03in} \noindent \bf {#1.\hspace{5pt}}}}

\def\BibTeX{{\rm B\kern-.05em{\sc i\kern-.025em b}\kern-.08em
    T\kern-.1667em\lower.7ex\hbox{E}\kern-.125emX}}
\begin{document}

\title{Out-of-Category Document Identification Using~Target-Category~Names as Weak Supervision}

\author{
\IEEEauthorblockN{Dongha Lee$^1$, Dongmin Hyun$^2$, Jiawei Han$^1$, Hwanjo Yu$^{2*}$}
\IEEEauthorblockA{
$^1$\textit{University of Illinois at Urbana-Champaign (UIUC), Urbana, IL, United States}\\
$^2$\textit{Pohang University of Science and Technology (POSTECH), Pohang, Republic of Korea}\\
\{donghal, hanj\}@illinois.edu, \{dm.hyun, hwanjoyu\}@postech.ac.kr}
\thanks{* Corresponding author}
}

\setlength{\abovecaptionskip}{5pt}
\setlength{\belowcaptionskip}{5pt}
\setlength{\floatsep}{7pt plus 1.0pt minus 2.0pt}
\setlength{\textfloatsep}{11pt plus 1.0pt minus 2.0pt}
\setlength{\dblfloatsep}{10pt plus 1.0pt minus 2.0pt}
\setlength{\dbltextfloatsep}{10pt plus 1.0pt minus 2.0pt}

\maketitle

\begin{abstract}
Identifying outlier documents, whose content is different from the majority of the documents in a corpus, has played an important role to manage a large text collection.
However, due to the absence of explicit information about the inlier (or target) distribution, existing unsupervised outlier detectors are likely to make unreliable results depending on the density or diversity of the outliers in the corpus. 
To address this challenge, we introduce a new task referred to as out-of-category detection, which aims to distinguish the documents according to their semantic relevance to the inlier (or target) categories by using the category names as weak supervision.
In practice, this task can be widely applicable in that it can flexibly designate the scope of target categories according to users' interests while requiring only the target-category names as minimum guidance.

In this paper, we present an out-of-category detection framework, which effectively measures how confidently each document belongs to one of the target categories based on its category-specific relevance score.
Our framework adopts a two-step approach;
(i) it first generates the pseudo-category label of all unlabeled documents by exploiting the word-document similarity encoded in a text embedding space, then (ii) it trains a neural classifier by using the pseudo-labels in order to compute the confidence from its target-category prediction.
The experiments on real-world datasets demonstrate that our framework achieves the best detection performance among all baseline methods in various scenarios specifying different target categories.
\end{abstract}


\section{Introduction}
\label{sec:intro}
Outlier detection for text data, which aims to identify semantically-deviating (or out-of-domain) documents from a large text corpus, has gained much attention for many real-world applications, such as automatic screening of patients' clinical records~\cite{hauskrecht2013outlier} and efficient management of websites or news articles~\cite{kannan2017outlier}. 
To handle the sparse and high-dimensional nature of text data, most existing work~\cite{yin2016model, kannan2017outlier, zhuang2017identifying, fouche2020mining, mohotti2020efficient} first learn a low-dimensional text embedding space or topic model of an input corpus in an unsupervised manner, then compute an \textit{outlier score} that indicates the outlierness of each document.
There have been various attempts to define an effective outlier score of each document, based on the local (or global) density of its embedding vector~\cite{breunig2000lof, sathe2016subspace}, its distance from semantic regions representing the normality~\cite{zhuang2017identifying, ruff2019self}, and the magnitude of its residuals in a term-document matrix~\cite{kannan2017outlier}.

The outlier detection task, however, can produce unreliable detection results, because the outlierness of a document is simply defined by its distance from the other documents without knowing the scope of the inlier categories (or topics).
To be specific, the existing detection methods tend to output high outlier scores for documents of an \textit{inlier-but-minor} category, because such inlier documents might differ from the majority of documents in an input corpus.
In addition, even the outlier documents clearly separable from the inliers become difficult to be detected if their proportion increases in the corpus.
In other words, this task is intrinsically error-prone due to the lack of explicit information (i.e., prior knowledge) about the inlier distribution.

\begin{figure}[t]
    \centering
    \includegraphics[width=\linewidth]{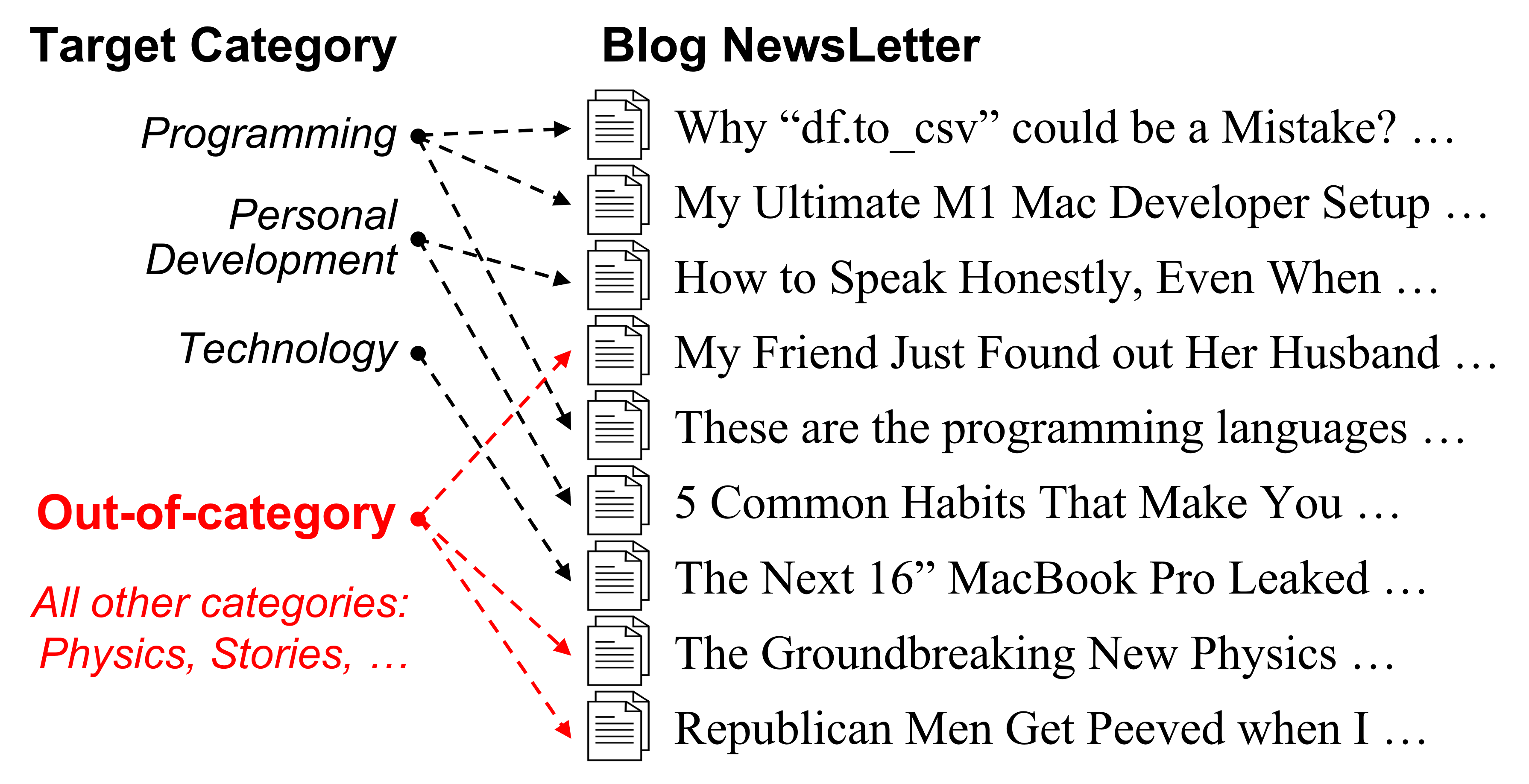}
    \caption{An illustrative example of our new task. It focuses on identifying \textit{out-of-category} documents that are not relevant to the given category names, rather than \textit{outlier} documents.}
    \label{fig:example}
\end{figure}

To tackle this challenge, we introduce a new task to leverage minimum guidance about inlier (or target) categories that a user has prior knowledge about or be interested in;
that is, the outlierness of each document can be clearly defined according to its semantic relevance to the categories.
Formally, given the set of target-category names (i.e., a single representative term for each category) for specifying an inlier (or target) distribution, our task focuses on distinguishing \textit{out-of-category} documents that do not belong to any of the target categories.
The target-category names can be regarded as \textit{weak supervision} in that it does not provide any document-level category information.
Note that this task does not require massive hand-labeled documents which are difficult to obtain, rather works on the set of unlabeled documents accompanied with the name of target categories.

Particularly, this out-of-category detection task is practical in that it is capable of flexibly designating target categories.
For example, Figure~\ref{fig:example} shows a corpus of blog newsletters that includes numerous articles.
Among a wide range of article topics, only three categories (\textit{Programming}, \textit{Personal Development}, and \textit{Technology}) are specified as the target.
Thus, the articles about the target categories are considered as in-category, while the rest of the articles should be identified as out-of-category, regardless of their category size or density in the corpus (\textit{Physics} and \textit{Stories}).
In this sense, conventional outliers which usually refer to out-of-domain documents can be treated as a special case of out-of-category, where all in-domain categories are specified as the target.

In this work, we present a novel framework for \underline{O}ut-\underline{O}f-\underline{C}ategory \underline{D}etection, termed as \proposed, which effectively computes the \textit{confidence} of all unlabeled documents by utilizing given target-category names as weak supervision.
Its main difference from unsupervised detectors is that it exploits the distinctiveness of target categories by the help of their names.
To identify the documents that do not belong to any of the target categories, our framework learns and utilizes category-indicative (i.e., discriminative) features that can determine the category membership of a document~\cite{hendrycks2020pretrained, lee2020multi}.
In detail, \proposed measures the confidence based on a two-step approach.
\begin{itemize}
    \item \textbf{Step 1: Embedding-based confidence.}
It first maps all words and documents into a joint text embedding space, assuming that they are category-conditionally generated from separable von Mises-Fisher (vMF) distributions.
Based on the obtained embedding space, it produces the pseudo-category labels of unlabeled documents to select confident documents that surely belong to one of the target categories.

    \item \textbf{Step 2: Classifier-based confidence.}
Using the set of confident documents and their pseudo-labels from the first step, it trains a neural classifier that accurately classifies an input document into target categories. 
In the end, \proposed computes the final confidence score of each document by using the output of the classifier, which is the maximum probability (or entropy) of the target-category prediction result.
\end{itemize}

Our extensive experiments on real-world datasets demonstrate that the proposed framework achieves the best performance for detecting outlier documents, and also for distinguishing out-of-category documents in a variety of scenarios with different target categories.
In particular, we qualitatively visualize our text embedding space, showing that it successfully encodes the discriminative category information of each document under the weak supervision.
For in-depth analyses on individual documents, we also investigate how the rank of their confidence varies depending on the types of detection methods.

\section{Related Work}
\label{sec:related}

\subsection{Outlier Text Detection}
\label{subsec:otd}
The goal of outlier text detection is to identify semantically-deviating (or out-of-domain) documents from a given text corpus.
The most dominant approach to this task is applying existing outlier detection methods on a low-dimensional vector space, where the semantic meaning of each document is effectively captured~\cite{mikolov2013distributed, meng2019spherical, zhuang2017identifying, fouche2020mining}.
Specifically, the outlierness of each document is computed by using the local outlier factor~\cite{breunig2000lof}, randomized hashing functions~\cite{sathe2016subspace}, or non-negative matrix factorization~\cite{kannan2017outlier}.
Recently, there have been several attempts to employ neural networks~\cite{ruff2019self, manolache2021date} for modeling the \textit{normality} of documents, regarding all unlabeled documents in the corpus as normal (i.e., inliers);
they detect the outliers based on how much each document deviates from the normality.

However, all the existing detection methods have critical limitations.
First, they only find out semantically minor documents in the corpus, without taking an actual underlying category (or topic) structure into consideration.
In this case, they might make incorrect predictions on the documents of \textit{inlier-but-minor} categories (i.e., false positive) or \textit{noisy-but-frequent} documents miscollected from other sources (i.e., false negative); 
this will be shown in our experiments (Section~\ref{subsec:qualanal}).
In other words, they cannot consider a set of inlier categories, which can be given as users' prior knowledge or interests.
For this reason, the scope of inliers and outliers needs to be designated more concretely, depending on the categories covered by the corpus.

Second, they do not explicitly learn the useful features related to inlier categories, which eventually results in limited performance for outlier detection.
They mainly utilize the embedding space optimized in an unsupervised manner;
the similarity among documents is implicitly captured by the co-occurrence of words, and this makes the documents not clearly distinguishable according to their category.
The most recent work on text classification~\cite{hendrycks2020pretrained, moon2021masker} empirically demonstrated that discriminative modeling among in-domain classes is helpful to enhance the robustness to the out-of-domain inputs.
Since class-indicative (i.e., discriminative) features are effective to determine whether an input belongs to a target class or not, they also can be used for detecting inputs that do not belong to any of the in-domain classes~\cite{lee2020multi}.
From this perspective, the unsupervised outlier detectors can be further improved by encouraging discrimination power for inlier categories.

\begin{figure*}[t]
    \centering
    \includegraphics[width=\linewidth]{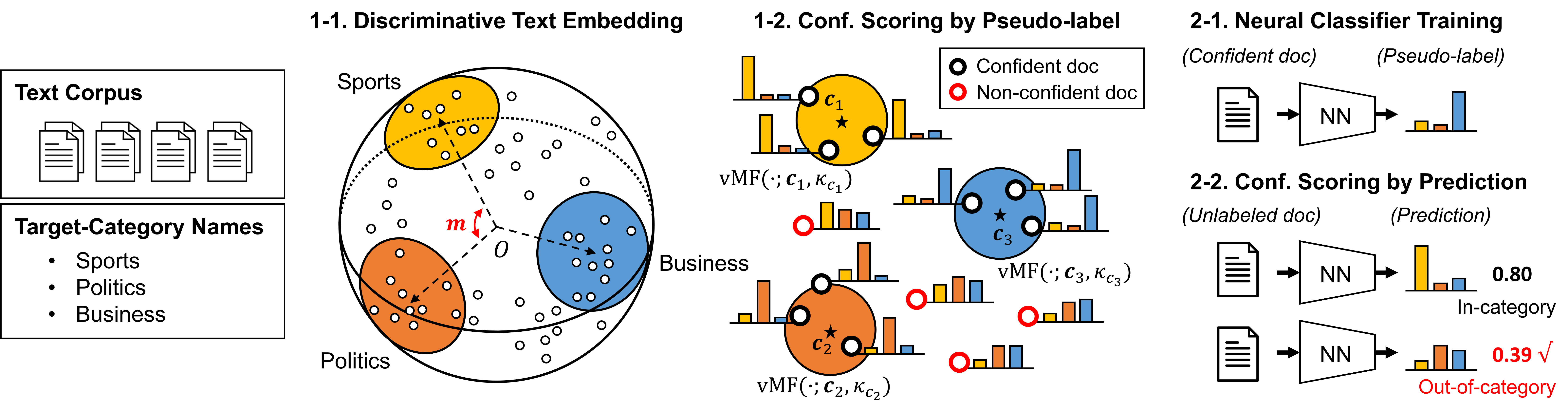}
    \caption{The overview of our \proposed framework for detecting out-of-category documents. It consists of two steps for scoring the confidence: (1) discriminative text embedding for generating the pseudo-category label of unlabeled documents, and (2) neural classifier training for making the target-category prediction on unlabeled documents.}
    \label{fig:framework}
\end{figure*}

\subsection{Weakly Supervised Text Classification}
\label{subsec:wsclf}
To alleviate the difficulty to obtain the class label of each document for text classification, several recent studies tried to train a text classifier by using unlabeled documents only with the label names or few keywords of each target class~\cite{meng2018weakly, meng2020text};
this task is called as \textit{weakly supervised} text classification.
The main challenge is to fully utilize various types of weak supervision for effectively training a text classifier.
To this end, existing methods adopt various techniques to infer the pseudo-label of unlabeled documents by distilling the knowledge from a pre-trained word embedding space~\cite{meng2018weakly, huang2020weakly}, or a pre-trained language model~\cite{meng2020text}. 

However, they assume that all unlabeled documents in a training corpus belong to one of the target classes, which implies that training and test documents are sampled from the in-domain distribution.
For this reason, although they are able to accurately classify in-domain documents into the target classes to some extent, they are not robust to out-of-domain documents.
To be specific, out-of-domain documents that reside in the training set can cause the classifier to make unreliable (i.e., high confident) predictions on out-of-domain inputs. 
To enhance the robustness by ensuring the ability to correctly identify out-of-domain documents, the training process needs to filter out non-confident documents that are less relevant to the target classes.

\section{Out-of-category Detection}
\label{sec:method}

\subsection{Problem Formulation}
\label{subsec:problem}
In this work, we focus on a weakly supervised outlier detection task with target-category names available, named as out-of-category detection.
Unlike unsupervised outlier detection, the goal of this task is to distinguish unlabeled documents according to their relevance (or similarity) to target categories.

\begin{definition}[Out-of-category document detection]
Given a set of unlabeled documents $\docuset=\{d_1, \ldots, d_N\}$ with their vocabulary $\wordset=\{w_1, \ldots, w_M\}$ and a set of target categories $\cateset=\{c_1,\ldots,c_K\}$ designated by their names $\wordset_\cateset=\{w_{c_1}, \ldots, w_{c_K}\}\subset\wordset$, we aim to obtain a measure of \textit{confidence}, denoted by $\text{conf}:\docuset \mapsto \mathbb{R}$, indicating how confidently each document belongs to the target categories.
\end{definition}

\subsection{Problem Analysis}
\label{subsec:analysis}
The key challenge of this problem is to model the confidence score of an unlabeled document based on its relevance to target categories.
The straightforward solutions for this challenge can be summarized into two approaches, depending on how to encode the category information of each document by utilizing the target-category names.

\smallsection{Using a text embedding space}
One possible solution is leveraging a joint embedding space of all words and documents~\cite{meng2019spherical, meng2020hierarchical}, to measure a document's relevance to each target category by the similarity of the document vector and the word (i.e., category name) vector in the embedding space.
This approach is effective to capture the category information of each document under weak supervision, since it additionally utilizes the co-occurrence between words and documents as self-supervision.
However, a text embedding space is not able to encode rich contextual information within a document.

\smallsection{Using a neural text classifier}
Another solution is training a target-category classifier which outputs the probability that an input document belongs to each category, based on a neural model with its capability of extracting useful features from a text.
The documents that contain one of the target-category names need to be collected to build the set of training documents (with the corresponding category labels),
but in this case, the classifier cannot be effectively trained due to the noisy labels and the limited number of labeled documents~\cite{meng2018weakly}.

\subsection{Overview}
\label{subsec:overview}
To get the best of both the approaches, our framework for out-of-category document detection, termed as \proposed, basically adopts a two-step approach that utilizes both a text embedding space and a neural classifier.
To be specific, \proposed aims to more effectively train the neural classifier by fully utilizing the knowledge encoded in the text embedding space, so that it can output the reliable confidence score based on its target-category prediction on unlabeled documents.
Figure~\ref{fig:framework} provides a high-level overview of our \proposed framework.

The first step, for embedding-based confidence scoring, maps all words, documents, and categories into a spherical embedding space, while making them discriminative based on given target-category names.
Using the embedding vectors, \proposed not only generates the pseudo-category label of all unlabeled documents, but also produces their confidence from the pseudo-label.
The second step, for classifier-based confidence scoring, trains a neural text classifier by using the set of confident documents and their pseudo-labels, filtered by the confidence in the first step.
Lastly, \proposed ranks all documents by their confidence, computed from the target-category prediction result.

\subsection{Embedding-based Confidence Scoring}
\label{subsec:confemb}
\subsubsection{Category-discriminative text embedding}
\label{subsubsec:josd}
To effectively capture the textual similarity (or distance) among words, documents, and categories into a joint embedding space, we employ the state-of-the-art spherical embedding framework~\cite{meng2019spherical} with an additional term using the target-category names for inter-category distinctiveness.
Due to the space limit, we briefly introduce the key idea and objective of our text embedding in this section.
Please refer to~\cite{meng2019spherical, meng2020hierarchical} for more details.

The objective of our text embedding is to maximize the generative likelihood of the corpus given the target categories $P(\docuset|\cateset)$, while enforcing that the category-conditional likelihood distributions are clearly separable.
In brief, based on the generative process, $P(\docuset|\cateset)$ is formulated to indicate how likely (i) each document $d_i$ comes from its category $c_{d_i}$, and (ii) each word $w_j$ co-occurs with its document $d_i$ and context words $w_k$. 
The loss function is described as follows.
\begin{equation}
\label{eq:josdloss}
\begin{split}
    \mathcal{L}_{emb} = -\log P(\docuset|\cateset) &+ \Omega(\cateset), \\
    P(\docuset|\cateset) = \prod_{d_i\in\docuset}p(d_i|c_{d_i}) &\prod_{w_j\in d_i} p(w_j|d_i) \hspace{-10pt} \prod_{w_k\in \text{cw}(w_j; d_i)} \hspace{-10pt} p(w_k|w_j) \\
    \approx \prod_{c_k\in\cateset} p(w_{c_k}|c_k) \cdot \prod_{d_i\in\docuset} &\prod_{w_j\in d_i} p(w_j|d_i) \hspace{-10pt} \prod_{w_k\in\text{cw}(w_j;d_i)} \hspace{-10pt} p(w_{k}|w_j),
\end{split}
\end{equation}
where $\Omega(\cateset)$ is the term for minimizing the semantic correlation between the target categories, defined by $\log \prod_{c_i,c_j\in\cateset} p(c_j|c_i)$, and $\text{cw}(w_j;d_i)$ is the set of surrounding words in a local context window for the center word $w_j$.
As the true category of each document $c_{d_i}$ is unknown, we replace the term $p(d_i|c_{d_i})$ with the category-conditional likelihood of target-category names $p(w_{c_k}|c_k)$ to utilize weak supervision. 

To optimize the embedding vector of each entity $w_j$, $d_i$, and $c_k$ (denoted by $\wvec{j}$, $\dvec{i}$, and $\cvec{k}$, respectively) based on $\mathcal{L}_{emb}$, we need to model each probability (or likelihood) in Equation~\eqref{eq:josdloss} by using the embedding vectors.
First of all, we define the generative likelihood of documents and words conditioned on each category, $p(d_i|c_k)$ and $p(w_j|c_k)$, by the von Mises-Fisher (vMF) distribution, which is a spherical distribution centered around $\cvec{k}$, to obtain a spherical space.
\begin{equation}
\label{eq:vmf}
\begin{split}
    p(d_i|c_k) &= \text{vMF}(\dvec{i};\cvec{k},\kappa_{c_k}) = n(\kappa_{c_k})\exp(\kappa_{c_k}\cos(\dvec{i}, \cvec{k})) \\
    p(w_j|c_k) &= \text{vMF}(\wvec{j};\cvec{k},\kappa_{c_k}) = n(\kappa_{c_k})\exp(\kappa_{c_k}\cos(\wvec{i}, \cvec{k}))
\end{split}
\end{equation}
where $\kappa_{c_k}\geq 0$ is the \textit{concentration} parameter, $n(\kappa_{c_k})$ is the normalization constant, and the \textit{mean direction} of each vMF distribution is modeled by the category embedding vector $\cvec{k}$.
Then, we also need to define the probability of word-document and word-word co-occurrence, $p(w_j|d_i)$ and $p(w_k|w_j)$, as well as that of inter-category correlation, $p(c_j|c_i)$. 
In this sense, we simply use the cosine (i.e., directional) similarity, which can be a measure of semantic coherence in the spherical space,
i.e., $p(w_j|d_i)\propto\exp(\cos(\wvec{j},\dvec{i}))$, $p(w_k|w_j)\propto\exp(\cos(\wvec{k},\wvec{j}))$, and $p(c_j|c_i)\propto\exp(\cos(\cvec{j},\cvec{i}))$.

Combining a max-margin loss function~\cite{vilnis2015word, vendrov2016order, ganea2018hyperbolic, meng2019spherical} with the category-conditional likelihood and the co-occurrence probability defined above, the objective of our text embedding in Equation~\eqref{eq:josdloss} is summarized as follows.
\begin{equation}
\label{eq:josdopt}
\begin{split}
    &\sum_{\substack{d_i\in\docuset}} \hspace{-25pt} \sum_{\substack{w_j\in d_i\\\hspace{26pt} w_k\in\text{cw}(w_j;d_i)}} \hspace{-25pt} \max \left(\vvec{k'}^\top\wvec{j} - \vvec{k}^\top\wvec{j} + \wvec{j'}^\top\dvec{i} - \wvec{j}^\top\dvec{i} + m, 0 \right) \\
    &\qquad - \sum_{c_k\in\cateset} \left(\log(n(\kappa_{c_k})) + \kappa_{c_k}\wvec{c_k}^\top\cvec{k}\right) \cdot \mathbbm{1}\left[\wvec{c_k}^\top\cvec{k} < m\right]\\
    &\qquad + \hspace{-4pt}\sum_{c_i,c_j\in\cateset}\max\left(\cvec{j}^\top\cvec{i} - m, 0 \right) \\
    &\quad \text{s.t.}\quad \forall w, d, c, \quad \lVert\wvec{ }\rVert=\lVert\vvec{ }\rVert=\lVert\dvec{ }\rVert=\lVert\cvec{ }\rVert=1, \kappa_c \geq 0,
\end{split}
\raisetag{38pt}
\end{equation}
where $m$ is the margin size and $\mathbbm{1}$ is the indicator function.
Similar to previous work on word embedding~\cite{mikolov2013distributed, pennington2014glove}, each word $w_j$ has two independent embedding vectors as the center word $\wvec{j}$ and the context word $\vvec{j}$, and the negative samples $w_{j'}$ and $w_{k'}$ are randomly selected from the vocabulary.

To sum up, the first term optimizes the similarity of each document and its words, and each word and its context words.
The second term pulls the category-indicative words (i.e., the given category names) close to the corresponding category vectors, while the third term makes the category vectors far apart from each other.

\subsubsection{Confidence scoring by target-category pseudo-labeling}
\label{subsubsec:docretrieval}
The category-conditional likelihood of a document $\text{vMF}(\dvec{};\cvec{},\kappa_c)$ in our text embedding space can serve as a good confidence measure, because it encodes the semantic relevance of documents to each target category.
Therefore, we generate soft pseudo-labels of all unlabeled documents by using their category-conditional likelihood, then calculate the confidence from the pseudo-label.

To this end, we present two strategies to measure the category-specific relevance score, $r:\docuset\times\cateset\mapsto \mathbb{R}$.
They either directly obtain the relevance by using a target document vector, denoted by $r_d(d, c)$, or indirectly capture it based on the proximity (i.e., nearest neighbor documents and words) in the embedding space, denoted by $r_w(d, c)$.
\begin{equation}
\label{eq:vmfscore}
\begin{split}
    r_{d}(d, c) &\propto \text{vMF}(\dvec{}; \cvec{}, \kappa_c) \\
    r_{w}(d, c) &\propto \hspace{-10pt} \sum_{(d', w)\in \mathcal{N}^{k,j}(d)} \hspace{-10pt} sim(\dvec{}, \dvec{}')\cdot sim(\dvec{}', \wvec{}) \cdot \text{vMF}(\wvec{}; \cvec{}, \kappa_c),
\end{split}
\end{equation}
where $\mathcal{N}^{k,j}(d)$ is the set of document-word pairs $(d', w)$ that consist of $k$ neighbor documents and their $j$ neighbor words identified by their similarity.
For the computation of $r_w$, we use the cosine similarity among documents and words, as all the embedding vectors reside on the spherical space.
This proximity-based relevance score can improve the robustness to the noise in the embedding space, by additionally leveraging the similarity of words and documents~\cite{fouche2020mining}.

Then, the pseudo-label of each document is obtained by normalizing the category-specific relevance scores over the target categories as follows.
\begin{equation}
\label{eq:pseudolabel}
    \hat{y}_c(d) = \frac{\exp( r(d, c)/T)}{\sum_{c'\in\cateset}\exp( r(d, c')/T)}
\end{equation}
where $T$ is the temperature parameter that controls the smoothness of a probability distribution~\cite{hinton2015distilling}.
Finally, the embedding-based confidence of document $d$ is defined by the maximum value of the soft pseudo-label $\hat{\mathbf{y}}(d)$.
\begin{equation}
\label{eq:confemb}
\text{conf}_{emb}(d) = \max_{c\in\mathcal{C}} \hat{y}_c(d) 
\end{equation}
This embedding-based confidence can be used for detecting out-of-category documents by itself, but \proposed makes use of it to filter out less confident documents for its next step.

\subsection{Classifier-based Confidence Scoring}
\label{subsec:confclf}
To take advantage of advanced neural architectures that effectively capture the contextual information within a document, \proposed utilizes a neural classifier for measuring the final confidence.
Note that the classifier takes the sequence of word tokens as an input document, not the embedding vector obtained from the first step.

\subsubsection{Neural classifier training}
\label{subsubsec:clftrain}
For training a neural classifier, we build the training set by utilizing both the pseudo-labels and confidences, described in Equation~\eqref{eq:pseudolabel} and~\eqref{eq:confemb}.
Depending on a given set of target categories, a large number of out-of-category documents could exist in the corpus, and using such documents for training the classifier degrades the performance of target-category discrimination.
Thus, we collect only the \textit{confident} documents, whose embedding-based confidence is larger than a filtering threshold $\tau_{emb}$, with their pseudo-labels as follows.
\begin{equation}
\label{eq:confdocuset}
\begin{split}
    &\mathcal{D}_{conf} = \left\{\left(d, \hat{\mathbf{y}}(d)\right) \vert  \text{ conf}_{emb}(d) > \tau_{emb}, \forall d \in \mathcal{D} \right\}
\end{split}
\end{equation}

\smallsection{Pre-training the classifier with pseudo-labels}
Using the confident documents in the training set, \proposed pre-trains a neural classifier by minimizing the cross-entropy between their pseudo-labels and the prediction output of the classifier.
This pre-training process distills the knowledge from the text embedding space into the classifier through the soft pseudo-labels~\cite{hinton2015distilling};
it eventually optimizes the classifier to copy the relevance of each confident documents to target categories.
\begin{equation}
\label{eq:ptloss}
    \ptloss = -\sum_{d\in\mathcal{D}_{conf}}\sum_{c\in\cateset} \hat{y}_c(d)\cdot\log p(c|d)
\end{equation}

\smallsection{Refining the classifier with self-training}
After the classifier is pre-trained by the confident documents and their pseudo-labels, \proposed further refines the classifier based on a self-training approach~\cite{xie2016unsupervised, meng2018weakly}.
The self-training process bootstraps the classifier;
its high-confident predictions on input documents are used to estimate their new targets (i.e., labels).
In detail, it gradually updates the output by minimizing the cross-entropy between the enhanced-but-consistent target $q(c|d)$ and the current prediction $p(c|d)$.
\begin{equation}
\label{eq:stloss}
    \stloss = -\sum_{d\in\mathcal{D}_{conf}}\sum_{c\in\cateset} q(c|d)\cdot\log p(c|d)
\end{equation}
The soft-label $q(c|d)$ is inferred by current prediction $p(c|d)$; 
i.e., $q(c|d)=\frac{p(c|d)^2/f(c)}{\sum_{c'\in\cateset}p(c'|d)^2/f(c')}$ where $f(c)=\sum_{d\in\mathcal{D}_{conf}}p(c|d)$ is the soft-frequency for category $c$.
Note that this is particularly effective for out-of-category detection, because it encourages to produce more confident prediction only for confident documents in the training set $\mathcal{D}_{conf}$.

\subsubsection{Confidence scoring by target-category prediction}
\label{subsubsec:oocd}
Similar to the embedding-based confidence computed from the pseudo-category label, the classifier-based confidence can be obtained from the category prediction result.
\proposed defines the final confidence by using the maximum softmax probability for target-category classification, which is the output of the neural classifier.
\begin{equation}
\label{eq:confclf}
    \text{conf}_{clf}(d) = \max_{c\in\cateset} p(c|d) = \max_{c\in\cateset} \frac{\exp(logit_{c, d})}{\sum_{c'\in\cateset}\exp(logit_{c',d})}
\end{equation}
Based on the final confidence, \proposed ranks all documents according to their confidence, which allows to distinguish out-of-category documents from in-category counterparts.

In addition to the maximum value of category prediction (i.e., maximum softmax probabiltiy), its negative entropy also can be used as a confidence measure~\cite{wang2019effective, fouche2020mining}; i.e., $\text{conf}_{clf}(d) = -\mathcal{H}[p(c|d)] = \sum_{c\in\cateset} p(c|d)\cdot\log p(c|d)$ where $\mathcal{H}$ is the entropy of an input distribution.
We empirically found that there is no significant difference between the two confidence measures in terms of detection performance, thus we simply use the maximum softmax probability.

\section{Experiments}
\label{sec:exp}

\begin{table}[t]
\caption{The statistics of the datasets.}
\centering
\resizebox{0.99\linewidth}{!}{%
\begin{tabular}{c|ccccc}
    \hline
    Corpus & Avg-Length & \#Categories & \#Inliers & \#Outliers  & Out-Ratio \\\hline
    \textbf{\nyt} & 725.8 & 26 & \ \ 13,081 & \ \ \ 113 & 0.0086 \\
    \textbf{\arxiv} & 127.3 & 34 & 169,172 & 3,326 & 0.0193 \\\hline
\end{tabular}
}
\label{tbl:datastats}
\end{table}

\subsection{Experimental Setting}
\label{subsec:expset}
\subsubsection{Datasets}
\label{subsubsec:dataset}

For our experiments, we use real-world corpora of two different domains:
\textbf{\nyt}\footnote{The news articles are crawled by using https://developer.nytimes.com/} and \textbf{\arxiv}\footnote{The abstracts of \arxiv papers are crawled from https://arxiv.org/}.
We crawled the documents from 26 categories in 5 different sections (for \nyt), and 34 categories in 3 different sections (for \arxiv).
To consider various types of outliers, we include the documents collected from the other categories/sections (i.e., local outlier) and the ones from the other domain (i.e., global outlier) while keeping their ratio very small (1$\sim$2\%).
The statistics and category information of each corpus is summarized in Tables~\ref{tbl:datastats} and~\ref{tbl:catinfo}.
For our problem setting that only target-category names are available, the category label of each document is not utilized at all for the task, but only for evaluation on each target scenario by determining whether each document is in-category or out-of-category.

\begin{table}[t]
\caption{All inlier category (and section) names.}
\centering
\resizebox{0.99\linewidth}{!}{%
\begin{tabular}{rll}
    \hline
    & \textbf{Section} & \textbf{Category} \\\hline
    \multirow{7}{*}{\rotatebox{90}{\nyt}}
    & politics & federal budget, surveillance, affordable care act, immigration, \\
    & & law enforcement, gay rights, gun control, military, abortion \\
    & arts & dance, television, music, movies \\
    & business & stocks and bonds, energy companies, economy, \\
    & & international business \\
    & science & cosmos, environment \\
    & sports & hockey, basketball, tennis, golf, football, baseball, soccer \\\hline
    \multirow{6}{*}{\rotatebox{90}{\arxiv}}
    & \multirow{1}{*}{math} & math.NA, math.AG, math.FA, math.NT, math.AP, math.OC, \\
    & & math.ST, math.PR, math.DG, math.CO, math.RT, math.DS,  \\
    & & math.GR, math.RA, math.SG, math.AT, math.MG \\
    & physics & ph.optics, ph.flu-dyn, ph.atom-ph, ph.ins-det, ph.acc-ph, \\
    & & ph.plasm-ph, ph.chem-ph, ph.class-ph \\
    & cs & cs.CV, cs.NI, cs.SE, cs.CC, cs.CR, cs.LO, cs.SY, cs.DS, cs.DB \\
    \hline
\end{tabular}
}
\label{tbl:catinfo}
\end{table}

For both the datasets, we make use of \autophrase~\cite{shang2018automated} to tokenize and segment raw texts of each document, thereby obtaining the phrase (or word) embedding vectors, as done in~\cite{fouche2020mining}.
Since our task only focuses on the documents in an input corpus, we do not consider other subword tokenizers~\cite{sennrich2016neural, kudo2018sentencepiece} mainly used to solve the out-of-vocabulary issue.

\subsubsection{Baselines}
\label{subsubsec:baseline}
We compare the performance of \proposed with that of other baseline methods which are designed for various tasks, including outlier detection, one-class classification, and weakly-supervised classification.
We re-categorize them as either (i) unsupervised methods that do not utilize the information about categories at all, and (ii) weakly supervised methods that can focus on target categories by utilizing the target category names.
The first category is unsupervised methods for text outlier detection.
\begin{itemize}
    \item \textbf{\acs}: A simple baseline that defines the outlier score of each document by its average negative cosine similarity to all the other documents in the corpus.
    
    \item \textbf{\lof}~\cite{breunig2000lof}: The most popular outlier detector based on the local density, which computes the local outlier factor.

    \item \textbf{\rshash}~\cite{sathe2016subspace}: A subspace hashing-based outlier detection method. We use the MurmurHash3 function with multiple random seeds for randomized hashing.
    
    \item \textbf{\tonmf}~\cite{kannan2017outlier}: A text outlier detector based on non-negative factorization of the term-document matrix. It computes the $l_2$-norm of each column (i.e., document) in its residual matrix.
    
    \item \textbf{\cvdd}~\cite{ruff2019self}: A neural one-class classifier designed for text data. It encodes an input document based on the multi-head self-attention architecture.
\end{itemize}
Since the density-based detection methods (i.e., \acs, \lof, and \rshash) work independently with the embedding space, we consider two spherical embedding spaces optimized with/without target-category names:
(i) the \textit{non-discriminative} space only capturing word-word and word-document contexts~\cite{meng2019spherical}, 
and (ii) the \textit{discriminative} space additionally enforcing discrimination of target categories (Section~\ref{subsec:confemb}).

The other category is weakly supervised methods that are capable of considering the semantic relevance between each unlabeled document and target categories to some extent.
\begin{itemize}
    \item \textbf{\vmfd}, \textbf{\vmfw}: The embedding-based confidence scoring methods that exploit separable vMF distributions (Equation~\eqref{eq:confemb}). \vmfd and \vmfw respectively adopt $r_d$ and $r_w$ as their category-specific relevance score (Equation~\eqref{eq:vmfscore}).\footnote{\vmfw can be thought as a weakly supervised variant of \kjnn~\cite{fouche2020mining}. For computing the category-specific relevance score (Equation~\eqref{eq:vmfscore}), \kjnn directly utilize the category label of each document, whereas \vmfw is tailored to use the discriminative embedding space obtained by the help of weak supervision.}
    
    \item \textbf{\cvddd}, \textbf{\cvddw}: The weakly supervised variants of \cvdd~\cite{ruff2019self}. Being tailored to use only the confident documents retrieved by \vmfd and \vmfw, they train the multi-head attention architecture for one-class classification.
    
    \item \textbf{\smclass}: A target-category classifier trained on a small labeled set, which includes only the documents that contain one of the category names (i.e., \underline{S}imple \underline{M}atch).
    
    \item \textbf{\westclass}~\cite{meng2018weakly}: A weakly supervised target-category classifier. It is trained on the set of pseudo-documents, whose words are generated by random sampling from each vMF distribution modeled in our embedding space. 
    
    \item \textbf{\proposedd}, \textbf{\proposedw}: The proposed confidence scoring based on our target-category classifier (Equation~\eqref{eq:confclf}). The training set of confident documents and their pseudo-labels is obtained by \vmfd and \vmfw, respectively.
    
\end{itemize}
All the methods based on target-category classification (i.e., \smclass, \westclass, \proposed) measure the confidence of unlabeled documents by using the classifier output; i.e., maximum softmax probability (Equation~\eqref{eq:confclf}).

\begin{table}[t]
\centering
\caption{Performance comparison for detecting outlier documents. (N) and (D) use the \underline{N}on-discriminative and \underline{D}iscriminative text embedding space, respectively. Best results are marked in bold and second best results are underlined.}
\resizebox{0.99\linewidth}{!}{%
\begin{tabular}{crcccccc}
\hline 
& & \multicolumn{3}{c}{\textbf{\nyt}} & \multicolumn{3}{c}{\textbf{\arxiv}}  \\ \cline{3-8}
& & {\small AUROC} & {\small AUPR} & {\small F1@O} & {\small AUROC} & {\small AUPR} & {\small F1@O} \\ \hline
\multirow{8}{*}{\rotatebox{90}{Unsupervised}}
& \acs (N) & 0.8704 & 0.1869 & 0.3328 & 0.7475 & 0.1407 & 0.2637 \\
& \acs (D) & 0.8734 & \underline{0.1919} & \underline{0.3451} & 0.7588 & 0.1752 & 0.2859 \\
& \lof (N) & 0.8477 & 0.1133 & 0.2566 & 0.7338 & 0.0951 & 0.1663 \\
& \lof (D) & 0.7884 & 0.0331 & 0.0265 & 0.7215 & 0.0629 & 0.0511 \\
& \rshash (N) & 0.8477 & 0.1133 & 0.2566 & 0.7538 & 0.0951 & 0.1663 \\
& \rshash (D) & 0.8689 & 0.1517 & 0.3009 & 0.7108 & 0.0719 & 0.1224 \\
& \tonmf & 0.5257 & 0.0097 & 0.0088 & 0.4892 & 0.0186 & 0.0177 \\
& \cvdd  & 0.8978 & 0.1574 & 0.0170 & 0.6277 & 0.0250 & 0.0353 \\
\hline
\multirow{8}{*}{\rotatebox{90}{Weakly supervised}}
& \vmfd & 0.7953 & 0.0269 & 0.0088 & \underline{0.8130} & \underline{0.1807} & \underline{0.3019} \\
& \vmfw & 0.8675 & 0.0419 & 0.0177 & 0.7339 & 0.0402 & 0.0244 \\
& \cvddd & 0.7849 & 0.1200 & 0.0170 & 0.6287 & 0.0252 & 0.0386 \\
& \cvddw & \underline{0.8989} & 0.1481 & 0.0170 & 0.7105 & 0.0416 & 0.0379 \\
& \smclass & 0.7050 & 0.0318 & 0.0265 & 0.5502 & 0.0199 & 0.0150 \\
& \westclass & 0.6064 & 0.0152 & 0.0088 & 0.4975 & 0.0184 & 0.0364 \\
& \proposedd  & \textbf{0.9399} & \textbf{0.3697} & 0.3628 & \textbf{0.8513} & \textbf{0.2695} & \textbf{0.3621} \\
& \proposedw  & 0.9091 & 0.3075 & \textbf{0.3894} & 0.7810 & 0.1643 & 0.1806 \\
\hline
\end{tabular}
}
\label{tbl:odperf}
\end{table}
\begin{table*}[thbp]
\caption{Four different evaluation scenarios for out-of-category document detection. Each scenario lists the target-category names with the ratio of out-of-category documents, whose category (or section) name is not included in the list.}
\centering
\resizebox{0.99\linewidth}{!}{%
\begin{tabular}{rlclc}
    \hline
    & \textbf{\nyt} & \textbf{Out-Ratio} & \textbf{\arxiv} & \textbf{Out-Ratio} \\\hline
    Major-Sec & sports, politics & 0.2353 & math, cs & 0.1779 \\
    Minor-Sec & science, business, arts & 0.7733 & cs, physics & 0.6991 \\
    Homo-Cat & hockey, tennis, basketball, golf & 0.7335 & math.(NA, AG, FA, NT, AP, OC) & 0.6868 \\
    Hetero-Cat & federal budget, music, stocks and bonds, environment, baseball & 0.7794 & math.(GR, RA, SG), ph.plasm-ph, cs.(CV, NI) & 0.8735 \\\hline
\end{tabular}
}
\label{tbl:targetinfo}
\end{table*}
\begin{table*}[thbp]
\caption{Performance comparison for detecting out-of-category documents. (N) and (D) use the \underline{N}on-discriminative and \underline{D}iscriminative text embedding space, respectively. Best results are marked in bold and second best results are underlined.}
\centering
\resizebox{0.99\linewidth}{!}{%
\begin{tabular}{ccrcccccccccccc}
\hline 
& & & \multicolumn{3}{c}{\textbf{Major-Section}} & \multicolumn{3}{c}{\textbf{Minor-Section}} & \multicolumn{3}{c}{\textbf{Homo-Category}} & \multicolumn{3}{c}{\textbf{Hetero-Category}} \\ \cline{4-15}
& & & {\small AUROC} & {\small AUPR} & {\small F1@O} & {\small AUROC} & {\small AUPR} & {\small F1@O} & {\small AUROC} & {\small AUPR} & {\small F1@O} & {\small AUROC} & {\small AUPR} & {\small F1@O} \\ \hline

\multirow{16}{*}{\rotatebox{90}{\textbf{\nyt}}} 
& \multirow{8}{*}{\rotatebox{90}{Unsupervised}}
& \acs (N) & 0.7920 & 0.4789 & 0.4781 & 0.2182 & 0.6483 & 0.7100 & 0.6890 & 0.8379 & 0.8002 & 0.4780 & 0.7698 & 0.7745 \\
& & \acs (D) & 0.7985 & 0.4947 & 0.4829 & 0.2150 & 0.6451 & 0.7103 & 0.6836 & 0.8331 & 0.8000 & 0.4935 & 0.7804 & 0.7772 \\
& & \lof (N) & 0.5118 & 0.2395 & 0.2490 & 0.5561 & 0.8150 & 0.7816 & 0.4801 & 0.7206 & 0.7284 & 0.5060 & 0.7865 & 0.7760 \\
& & \lof (D) & 0.5008 & 0.2316 & 0.2291 & 0.5697 & 0.8214 & 0.7840 & 0.4714 & 0.7124 & 0.7271 & 0.5125 & 0.7877 & 0.7785 \\
& & \rshash (N) & 0.7889 & 0.4631 & 0.4646 & 0.2204 & 0.6508 & 0.7080 & 0.5835 & 0.7827 & 0.7605 & 0.6633 & 0.8334 & 0.8442 \\
& & \rshash (D) & 0.7776 & 0.4541 & 0.4617 & 0.2307 & 0.6562 & 0.7102 & 0.6220 & 0.8035 & 0.7753 & 0.6516 & 0.8360 & 0.8327 \\
& & \tonmf & 0.4933 & 0.2335 & 0.2423 & 0.5081 & 0.7762 & 0.7711 & 0.5049 & 0.7355 & 0.7313 & 0.4900 & 0.7726 & 0.7840 \\
& & \cvdd  & 0.5796 & 0.2610 & 0.1295 & 0.4381 & 0.7756 & 0.7538 & 0.5858 & 0.7558 & 0.7259 & 0.4804 & 0.7662 & 0.7969 \\
\cline{2-15}
& \multirow{8}{*}{\rotatebox{90}{Weakly supervised}}
& \vmfd & 0.7607 & 0.4811 & 0.5071 & 0.9101 & 0.9629 & 0.9305 & 0.9570 & 0.9631 & 0.9349 & 0.8799 & 0.9660 & 0.8795 \\
& & \vmfw & \underline{0.8121} & \underline{0.5295} & \underline{0.5620} & \underline{0.9629} & \underline{0.9849} & \underline{0.9499} & \underline{0.9711} & \underline{0.9846} & \underline{0.9509} & \underline{0.8960} & \underline{0.9689} & \underline{0.9055} \\
& & \cvddd & 0.6718 & 0.3427 & 0.0744 & 0.3376 & 0.7069 & 0.8039 & 0.6291 & 0.7964 & 0.7027 & 0.5399 & 0.8052 & 0.7665 \\
& & \cvddw & 0.6055 & 0.2730 & 0.1140 & 0.4118 & 0.7660 & 0.7653 & 0.5890 & 0.7510 & 0.7199 & 0.4949 & 0.7748 & 0.7962 \\
& & \smclass & 0.7374 & 0.3747 & 0.4037 & 0.7240 & 0.8915 & 0.8359 & 0.7431 & 0.8379 & 0.8373 & 0.7392 & 0.8822 & 0.8471 \\
& & \westclass & 0.5236 & 0.2502 & 0.2619 & 0.5405 & 0.7941 & 0.7831 & 0.5249 & 0.7520 & 0.7364 & 0.4691 & 0.7615 & 0.7734 \\
& & \proposedd  & 0.9318 & \textbf{0.8245} & \textbf{0.7845} & 0.9387 & 0.9793 & 0.9353 & 0.9667 & 0.9866 & 0.9424 & \textbf{0.9436} & \textbf{0.9804} & \textbf{0.9434} \\
& & \proposedw  & \textbf{0.9445} & 0.7384 & 0.7726 & \textbf{0.9776} & \textbf{0.9928} & \textbf{0.9585} & \textbf{0.9842} & \textbf{0.9934} & \textbf{0.9672} & 0.9360 & 0.9743 & 0.9409 \\

\hline
\multirow{16}{*}{\rotatebox{90}{\textbf{\arxiv}}} 
& \multirow{8}{*}{\rotatebox{90}{Unsupervised}}
& \acs (N) & 0.4566 & 0.1526 & 0.1104 & 0.1621 & 0.5272 & 0.5941 & 0.6575 & 0.8180 & 0.7363 & 0.3616 & 0.8266 & 0.8670 \\
& & \acs (D) & 0.5437 & 0.1888 & 0.1627 & 0.3455 & 0.6048 & 0.6429 & 0.6436 & 0.8051 & 0.7354 & 0.3643 & 0.8244 & 0.8674 \\
& & \lof (N) & 0.4951 & 0.1751 & 0.1717 & 0.5526 & 0.7275 & 0.7276 & 0.5378 & 0.7094 & 0.7022 & 0.6142 & 0.9040 & 0.8893 \\
& & \lof (D) & 0.5280 & 0.2026 & 0.2199 & 0.5680 & 0.7248 & 0.7404 & 0.5477 & 0.7191 & 0.7057 & 0.6228 & 0.906  & 0.8905 \\
& & \rshash (N) & 0.5282 & 0.1817 & 0.1774 & 0.3112 & 0.6088 & 0.6198 & 0.6439 & 0.7744 & 0.7510 & 0.4543 & 0.8654 & 0.8679 \\
& & \rshash (D) & 0.5155 & 0.1777 & 0.1674 & 0.4794 & 0.7016 & 0.6833 & 0.5376 & 0.7127 & 0.7009 & 0.4704 & 0.8677 & 0.8692 \\
& & \tonmf & 0.4976 & 0.1766 & 0.1805 & 0.5007 & 0.7000 & 0.6993 & 0.4995 & 0.6865 & 0.6859 & 0.5007 & 0.8735 & 0.8734 \\
& & \cvdd  & 0.6071 & 0.2236 & 0.1082 & 0.4289 & 0.6775 & 0.7191 & 0.5491 & 0.7029 & 0.6769 & 0.4596 & 0.8667 & 0.8731 \\
\cline{2-15}
& \multirow{8}{*}{\rotatebox{90}{Weakly supervised}}
& \vmfd & 0.6429 & 0.2416 & 0.2512 & 0.7561 & 0.8310 & 0.8219 & 0.6866 & \underline{0.8016} & 0.7725 & \underline{0.8285} & \underline{0.9637} & \underline{0.9227} \\
& & \vmfw & 0.4634 & 0.1530 & 0.1055 & \underline{0.8741} & \underline{0.9058} & \underline{0.8853} & \underline{0.6965} & 0.7981 & \underline{0.7774} & 0.8007 & 0.9555 & 0.9162 \\
& & \cvddd & 0.4997 & 0.1681 & 0.1590 & 0.4663 & 0.6955 & 0.7186 & 0.5519 & 0.7069 & 0.6769 & 0.5415 & 0.8880 & 0.8696 \\
& & \cvddw & 0.5426 & 0.1877 & 0.1519 & 0.6549 & 0.7992 & 0.6339 & 0.5936 & 0.7320 & 0.6661 & 0.5109 & 0.8803 & 0.8722 \\
& & \smclass & \underline{0.7225} & \underline{0.3059} & \underline{0.3206} & 0.6236 & 0.7391 & 0.7659 & 0.5918 & 0.7343 & 0.7327 & 0.7257 & 0.9318 & 0.9178 \\
& & \westclass & 0.5541 & 0.2236 & 0.2511 & 0.5652 & 0.7222 & 0.7329 & 0.4589 & 0.6765 & 0.6619 & 0.5483 & 0.8891 & 0.8769 \\
& & \proposedd  & 0.8127 & 0.4165 & 0.4543 & 0.7868 & 0.8625 & 0.8353 & \textbf{0.7393} & \textbf{0.8373} & \textbf{0.7989} & \textbf{0.8421} & \textbf{0.9742} & \textbf{0.9310} \\
& & \proposedw  & \textbf{0.8514} & \textbf{0.4973} & \textbf{0.5294} & \textbf{0.8919} & \textbf{0.9286} & \textbf{0.9067} & 0.7127 & 0.8107 & 0.7906 & 0.8309 & 0.9652 & 0.9251 \\
\hline
\end{tabular}
}
\label{tbl:oocdperf}
\end{table*}

\subsubsection{Evaluation metrics}
\label{subsubsec:baseline}
As evaluation metrics for our detection tasks, we measure (i) the area under the receiver operating curve (\textbf{AUROC}), (ii) the area under the precision-recall curve (\textbf{AUPR}),\footnote{In cases of AUPR and F1, we measure the values where out-of-category (or outlier) documents are considered as positive.} 
and (iii) the F1 score at a top-$O$ list of documents (\textbf{F1@O}), where $O$ is the number of actual out-of-category (or outlier) documents; this is equivalent to using the confidence threshold $\Gamma$ that satisfies $p_{out}\cdot{|\docuset|} == {|\{d| \text{conf}(d) <\Gamma, \forall d\in\docuset \}|}$, where $p_{out}$ is the out-of-category (or outlier) ratio for each target scenario.
For the classifier-based methods, including \cvdd, \smclass, \westclass, and \proposed, we report the average of three independent runs, each of which uses different random seeds for initialization.

\subsubsection{Implementation details}
We implement our \proposed framework and the other baseline methods by using PyTorch,\footnote{All the experiments are conducted on NVIDIA Titan Xp.} except for using the official author codes of \tonmf\footnote{https://github.com/ramkikannan/outliernmf} and \cvdd\footnote{https://github.com/lukasruff/CVDD-PyTorch}.
For a fair comparison, \smclass, \westclass, and \proposed adopt the same CNN architecture with a single 1D convolutional layer~\cite{kim-2014-convolutional}.
We initialize their word embedding layer by the embedding vectors obtained from our first step, and use the Adam optimizer to train each classifier.
We simply fix the temperature parameter $T$ to 0.1,\footnote{We empirically found that this hyperparameter hardly affects the final performance, and the sensitivity analysis will be provided in Section~\ref{subsec:paramanal}.} while tuning the filtering threshold $\tau_{emb}$ with respect to the ratio of confident documents in the training set,
$|\mathcal{D}_{conf}|/|\mathcal{D}|\in\{0.1, \ldots, 1.0\}$.
In cases of \vmfw, \cvddw, and \proposedw, the numbers of neighbor documents and words in their relevance scores are set to the values suggested by~\cite{fouche2020mining}, i.e., $k=j=30$.

\subsection{Outlier detection}
\label{subsec:perf_od}
We first evaluate all the methods in terms of identifying a small number of outlier documents among a large number of unlabeled documents in a text corpus.
In Table~\ref{tbl:odperf}, our frameworks (i.e., \proposedd and \proposedw) achieve the best performance for both the datasets.
We observe that there is no remarkable difference between unsupervised methods and weakly supervised methods, except for the \proposed framework.
This is because the outliers, which are different from the majority of the inliers, can be detected even in an unsupervised way to some extent, by using their low density or the deviation from the normality.
On the contrary, \proposed leverages prior knowledge about the scope of the inlier categories in the corpus so that it measures the confidence of unlabeled documents by their semantic relevance to the inlier categories.
This allows to make reliable results not being affected by the density or diversity of the outliers, which leads to significant improvement of the outlier detection performance.

\subsection{Out-of-category Detection}
\label{subsec:perf_ooc}

Next, we compare the out-of-category detection performance of \proposed with that of the other baselines.
We consider four different scenarios where target categories are flexibly designated as listed in Table~\ref{tbl:targetinfo}.
\begin{itemize}
    \item \textbf{Major-Section}, \textbf{Minor-Section}:
    To demonstrate that high-level category (i.e., section) names also can be used for detecting out-of-category documents, we choose the subsets of sections by their number of documents in descending (i.e., major) and ascending (i.e., minor) order.
    
    \item \textbf{Homo-Category}, \textbf{Hetero-Category}:
    To consider different levels of semantic correlation among target categories, we select the categories from a single section (i.e., homogeneous) or multiple sections (i.e., heterogeneous).
\end{itemize}
In case of \arxiv, we use a single main keyword of each category instead of its category name, since each category name is an abbreviation that is difficult to be correctly embedded into the text embedding space.

Table~\ref{tbl:oocdperf} shows that \proposed considerably outperforms the other baselines for all the target scenarios. 
To be specific, the unsupervised outlier detection methods fail to distinguish the out-of-category documents.
Particularly, they show poor performance as the ratio of out-of-category documents increases in the corpus (e.g., Minor-Section and Hetero-Category), because they mainly employ the similarity (or distance) to other documents rather than to the target categories.

Among the weakly supervised methods, \proposed beats each type of the baselines in the following aspects:

\smallsection{Comparison with the embedding-based methods (\vmfd and \vmfw)}
Compared to the confidence scoring methods based on the document embedding vectors, a neural classifier of \proposed is better at capturing the contextual information in each document, which eventually helps to accurately compute the category-specific semantic relevance scores.

\smallsection{Comparison with the one-class classifiers (\cvddd and \cvddw)}
\proposed learns category-discriminative features of documents by training a multi-class classifier, so it can leverage much richer category (or topic) information than \cvdd which learns stereotypical (i.e., normal) features of in-category documents.
As a consequence, a target-category classifier more effectively distinguishes the out-of-category documents compared to a one-class classifier.

\smallsection{Comparison with the target-category classifiers (\smclass and \westclass)}
The existing weakly supervised classifiers show poor performances in spite of their neural architecture.
This result strongly indicates that pseudo-labeling the documents on the embedding space is a more effective way to distill the knowledge from the embedding space, compared to simply using the limited number of exactly matched documents (\smclass) or synthesized pseudo-documents (\westclass).

\begin{table*}[t]
\centering
\caption{The AUROC for detecting out-of-category (or outlier) documents, while ablating each component of \proposedd.}
\resizebox{0.99\linewidth}{!}{%
\begin{tabular}{cccccccccccccc}
\hline 
\textbf{Confidence} & \textbf{Filtering} & \textbf{Temperature} & \textbf{Self-Training} & \multicolumn{2}{c}{\textbf{Inlier-Category}} & \multicolumn{2}{c}{\textbf{Major-Section}} & \multicolumn{2}{c}{\textbf{Minor-Section}} & \multicolumn{2}{c}{\textbf{Homo-Category}} & \multicolumn{2}{c}{\textbf{Hetero-Category}} \\ 

\textbf{Scoring} & $\tau_{emb}$ & $T$ & $\stloss$ & \nyt & \arxiv & \nyt & \arxiv & \nyt & \arxiv & \nyt & \arxiv & \nyt & \arxiv \\ \hline
\multirow{2}{*}{$\text{conf}_{emb}$} 
& & & & 0.7953 & 0.8130 & 0.7607 & 0.6429 & 0.9101 & 0.7561 & 0.9570 & 0.6866 & 0.8799 & 0.8285 \\
& & \checkmark& & 0.8004 & 0.8173 & 0.7607 & 0.6429 & 0.9274 & 0.7761 & 0.9697 & 0.7126 & 0.8950 & 0.8399 \\\hline

\multirow{5}{*}{$\text{conf}_{clf}$}
& & \checkmark & \checkmark 
& 0.8978 & 0.8185 & 0.8493 & 0.7140 & 0.8960 & 0.7224 & 0.9322 & 0.6605 & 0.9164 & 0.7689 \\
& \checkmark & & & 0.8890 & 0.8035 & 0.8627 & 0.7576 & 0.9127 & 0.7686 & 0.9669 & 0.6773 & 0.8873 & 0.8189 \\
& \checkmark & & \checkmark & 0.9246 & 0.8362 & 0.9214 & 0.7984 & 0.9246 & 0.7779 & \textbf{0.9767} & 0.7150 & 0.9219 & 0.8310  \\
& \checkmark & \checkmark & & 0.9134 & 0.8242 & 0.8724 & 0.7861 & 0.9286 & 0.7734 & 0.9551 & 0.7058 & 0.8913 & 0.8318 \\
& \checkmark & \checkmark & \checkmark & \textbf{0.9399} & \textbf{0.8513} & \textbf{0.9318} & \textbf{0.8127} & \textbf{0.9387} & \textbf{0.7868} & 0.9667 & \textbf{0.7393} & \textbf{0.9436} & \textbf{0.8421} \\
\hline
\end{tabular}
}
\label{tbl:ablation}
\end{table*}


\begin{table*}[t]
\centering
\caption{The confidence rank of each document in the \nyt corpus, obtained by each detection method.}
\resizebox{0.99\linewidth}{!}{%
\begin{tabular}{|X|B|B|B|B|}
\hline
\textbf{Target} & \textbf{all \nyt inlier categories} & \textbf{all \nyt inlier categories} & \textbf{all \nyt inlier categories} & \textbf{hockey}, \textbf{tennis}, \textbf{basketball}, \textbf{golf} \\
\hline
\textbf{Document}
&
\multicolumn{1}{L|}{Two NASA astronauts, working more quickly than expected, completed a spacewalk on Saturday in which they took the first steps to repair a malfunctioning pump module that is part of the cooling system for the International Space Station. The astronauts, Col. Michael S. Hopkins of the Air Force and Richard A. Mastracchio, were far ahead of schedule [...]}
&
\multicolumn{1}{L|}{DANA HAMEL, a local interior designer, works an average of 10 hours a day, eats in restaurants four nights a week and spends three months every year traveling to California on business. So when Mr. Hamel, who is 34 and originally from Santa Monica, Calif., heard about a new building with hotel rooms, condos and stores going up in the South Lake Union [...]}
&
\multicolumn{1}{L|}{Shape optimization based on the shape calculus is numerically mostly performed by means of steepest descent methods. This paper provides a novel framework to analyze shape-Newton optimization methods by exploiting a Riemannian perspective. A Riemannian shape Hessian is defined yielding often sought properties like symmetry and quadratic convergence [...]}
&
\multicolumn{1}{L|}{KANSAS CITY, Mo. — Christian Yelich singled home the go-ahead run with one out in the 10th inning Tuesday night and the Miami Marlins beat the Kansas City Royals 1-0 after a tidy matchup of contrasting starters. Hard-throwing Marlins prodigy Jose Fernandez and wily Royals veteran Bruce Chen each lasted seven innings before handing the scoreless game [...]} \\
\hline
\textbf{Category} & \textbf{cosmos} & \textbf{real estate} & \textbf{math.OC} & \textbf{baseball} \\ 
\hline
\textbf{In/Out} & \textbf{in-category} & \textbf{out-of-category} & \textbf{out-of-category} & \textbf{out-of-category} \\
\hline
\multirow{4}{*}{\textbf{ConfRank}}
& [\makebox[0.6in][r]{\acs}]\quad Top-88\% \makebox[0.3in][c]{\xmark} & [\makebox[0.6in][r]{\acs}]\quad Top-35\% \makebox[0.3in][c]{\xmark} & [\makebox[0.6in][r]{\acs}]\quad Top-76\% \makebox[0.3in][c]{\cmark} & [\makebox[0.6in][r]{\acs}]\quad Top-25\% \makebox[0.3in][c]{\xmark} \\ 
& [\makebox[0.6in][r]{\lof}]\quad Top-90\% \makebox[0.3in][c]{\xmark} & [\makebox[0.6in][r]{\lof}]\quad Top-71\% \makebox[0.3in][c]{\cmark} & [\makebox[0.6in][r]{\lof}]\quad Top-03\% \makebox[0.3in][c]{\xmark} & [\makebox[0.6in][r]{\lof}]\quad Top-11\% \makebox[0.3in][c]{\xmark} \\ 
& [\makebox[0.6in][r]{\vmfd}]\quad Top-28\% \makebox[0.3in][c]{\cmark} & [\makebox[0.6in][r]{\vmfd}]\quad Top-11\% \makebox[0.3in][c]{\xmark} & [\makebox[0.6in][r]{\vmfd}]\quad Top-97\% \makebox[0.3in][c]{\cmark} & [\makebox[0.6in][r]{\vmfd}]\quad Top-20\% \makebox[0.3in][c]{\xmark} \\ 
& [\makebox[0.6in][r]{\proposedd}]\quad Top-15\% \makebox[0.3in][c]{\cmark} & [\makebox[0.6in][r]{\proposedd}]\quad Top-69\% \makebox[0.3in][c]{\cmark} & [\makebox[0.6in][r]{\proposedd}]\quad Top-91\% \makebox[0.3in][c]{\cmark} & [\makebox[0.6in][r]{\proposedd}]\quad Top-86\% \makebox[0.3in][c]{\cmark} \\ \hline
\end{tabular}
}
\label{tbl:casestudy}
\end{table*}

\begin{figure}[thbp]
    \centering
    \includegraphics[width=\linewidth]{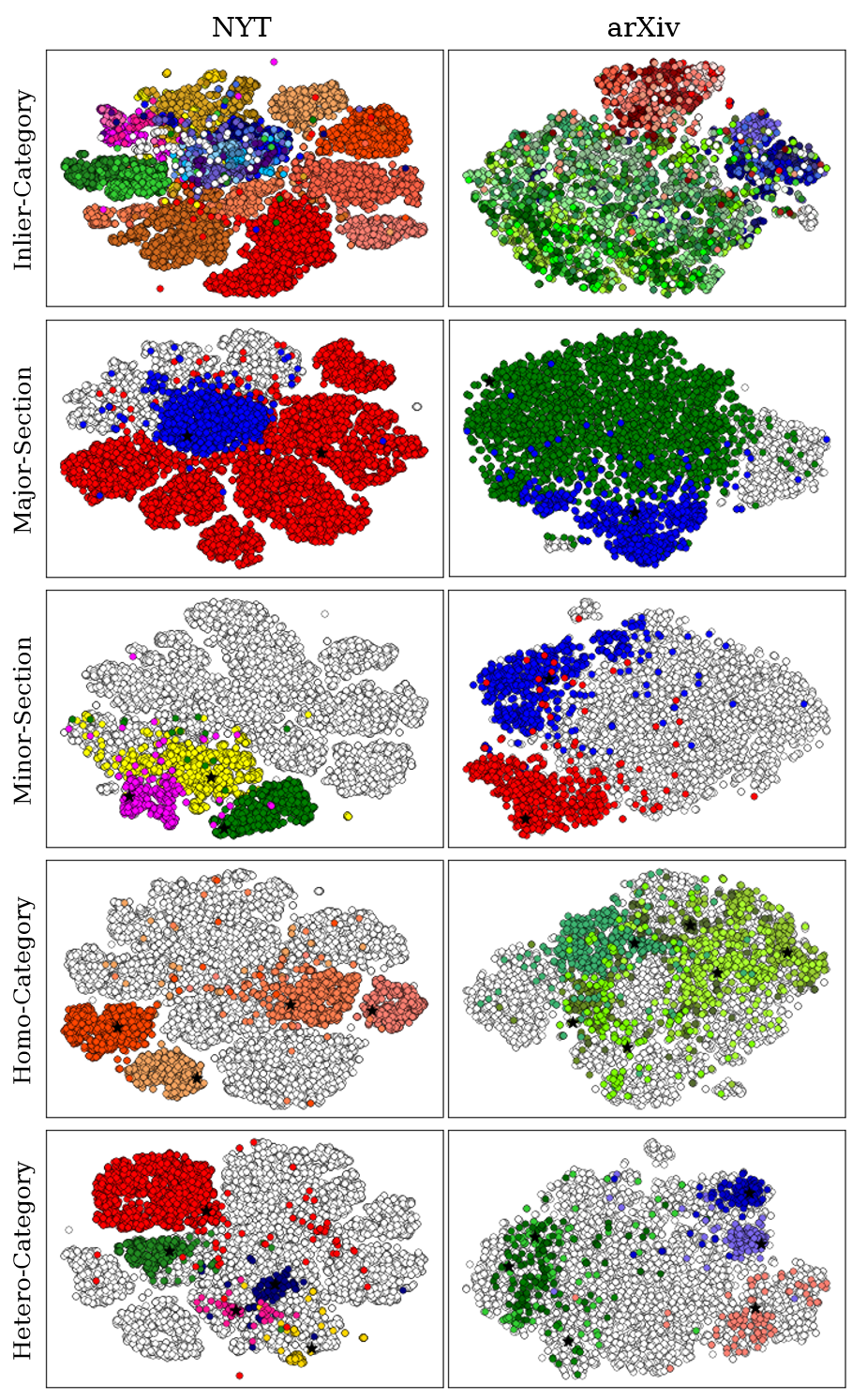}
    \caption{The visualization of the discriminative text embedding space. Colored and white circles represent in-category and out-of-category documents, respectively, and black asterisks show the category embedding vectors (i.e., the mean direction of each vMF distribution). Best viewed in color.}
    \label{fig:embspace}
\end{figure}

\subsection{Ablation Study}
\label{subsec:ablation}
We provide an ablation analysis on out-of-category detection performance, to validate the effectiveness of the following components:
(i) a two-step approach to confidence scoring $\text{conf}_{clf}$, 
(ii) training the classifier on confident documents retrieved by a filtering threshold $\tau_{emb}$, 
(iii) pseudo-labeling by softmax normalization with the temperature parameter $T$, and 
(iv) refining the classifier via self-training process $\stloss$. 

Table~\ref{tbl:ablation} reports the AUROC on both the datasets in various scenarios.\footnote{Inlier-Category is the scenario that all inlier category names are specified.} 
As discussed in Section~\ref{subsec:perf_ooc}, $\text{conf}_{clf}$ consistently shows higher AUROC scores than $\text{conf}_{emb}$ with the assistance of its neural classifier.
However, in case of using the entire corpus to train the classifier (without the filtering threshold $\tau_{emb}$), $\text{conf}_{clf}$ does not work well because the target-category classifier is trained by numerous out-of-category documents with their noisy labels.
Furthermore, both the temperature $T$ and the self-training loss $\stloss$ are helpful to enhance the detection performance of our \proposed framework, by making it produce more confident (i.e., sharper) pseudo-category labels and target-category prediction, respectively.
In particular, $\stloss$ significantly increases the AUROC compared to the case of using only $\ptloss$ for all the target scenarios.

\subsection{Qualitative Analysis}
\label{subsec:qualanal}

We visualize the discriminative embedding space for each target scenario by using t-SNE.
In case of \arxiv that has a large number of documents, we plot 10,000 document vectors randomly selected from the corpus.
Each document vector is colored according to its true category, where each color group is used to mark different sections so as to indicate the semantic correlation among the target categories.

Figure~\ref{fig:embspace} shows that document embedding vectors gather around each category embedding vector to follow the vMF distribution, thus they can be distinguished in our embedding space according to their categories to some extent.
In case of \arxiv, some categories fail to be aligned with their documents, because the category information of the \arxiv documents is difficult to be accurately captured by only the co-occurrence of words and documents.
Nevertheless, the similarity (or distance) of document vectors to each category vector successfully encodes their semantic relevance to the category in general, thus it can serve as a measure of confidence used to distinguish out-of-category from in-category documents.

In addition, we examine how the confidence rank of each document varies depending on detection methods: \acs, \lof, \vmfd, and \proposedd.
Table~\ref{tbl:casestudy} describes the result on four example documents from the \nyt dataset.
In summary, \proposedd correctly ranks all of them while the other methods fail.
The first document, which belongs to an inlier-but-minor category (\textit{cosmos}), is correctly identified as inlier by \vmfd and \proposedd, because the weakly supervised methods are aware that \textit{cosmos} is one of the inlier categories and utilize its relevance to the category.
In cases of the second/third documents, which are the representatives of local/global outliers (\textit{real estate} and \textit{math.OC}), each of them is incorrectly ranked by \acs and \lof, respectively.
To be precise, the unsupervised detection methods are likely to make unreliable predictions when these outlier documents become locally or globally dense, as discussed in Section~\ref{subsec:otd}.
Finally, the fourth document from out-of-category (\textit{baseball}) is only correctly identified by \proposed among the weakly supervised methods.
Its relevance to some of the target categories (e.g., \textit{tennis} and \textit{basketball}) can be overestimated by the embedding-based method \vmfd, due to their similar word occurrence, such as \textit{game}, \textit{matchup}, and \textit{scoreless}.

\subsection{Parameter Analysis}
\label{subsec:paramanal}
Finally, we study how sensitive the performance of \proposed is to its hyperparameters: 
(i) the filtering threshold $\tau_{emb}$ to build the set of confident documents, and
(ii) the temperature parameter $T$ for pseudo-labeling.
We first assess the quality of $\docuset_{conf}$ in terms of the pseudo-label consistency, defined by $\frac{1}{|\docuset_{conf}|}\cdot\sum_{(d, \hat{\mathbf{y}}(d))\in\docuset_{conf}}\mathbbm{1}[\argmax_{c\in\cateset}\hat{y}_c(d)==y(d)]$, varying the filtering threshold $\tau_{emb}$ .
Then, we investigate the final performance of \proposedd with respect to the hyperparameters.

\begin{figure}[t]
	\centering
	\includegraphics[width=\linewidth]{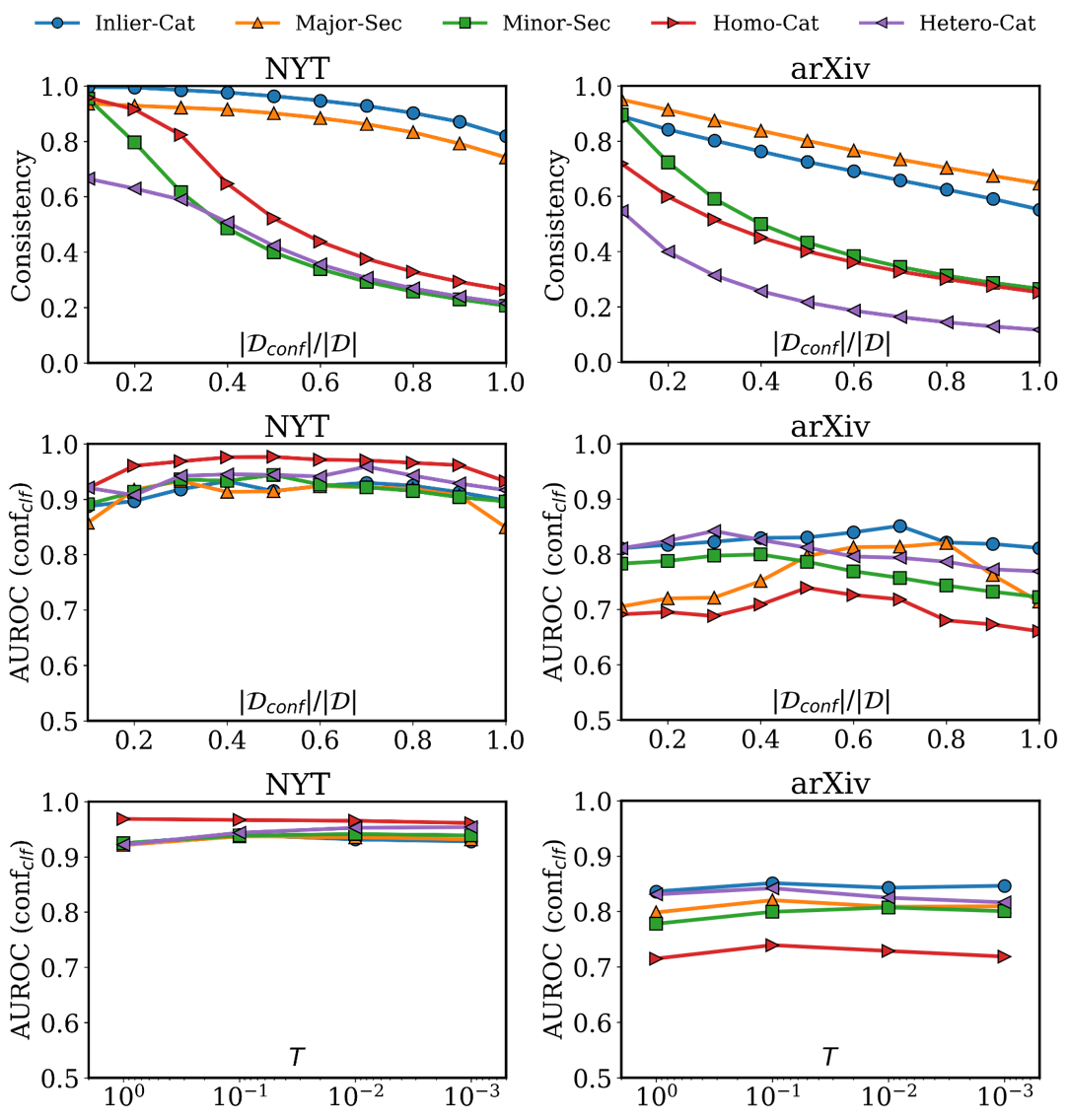}
	\caption{The performance of \proposedd varying $\tau_{emb}$ and $T$.}
	\label{fig:paramanal}
\end{figure}

In Figure~\ref{fig:paramanal}, the pseudo-label consistency gets lower as the size of $|\docuset_{conf}|$ becomes larger, showing that $\tau_{emb}$ controls the trade-off between the number of confident documents and the accuracy of their pseudo-labels.
For this reason, the final performance of \proposedd becomes worse in both the cases of using only a very small number of surely-confident documents or simply using all documents regardless of their confidence.
On the other hand, the performance does not largely depend on the choice of $T$, even though its smaller value brings the improvement compared to the standard softmax normalization (i.e., $T=1$).
In conclusion, \proposed can achieve better discrimination between in-category and out-of-category documents with the help of the proper hyperparameter values.

\section{Conclusion}
\label{sec:conc}
This paper proposes a new task for detecting out-of-category documents from a text corpus, by using given target-category names as weak supervision. 
To effectively measure the semantic relevance between each document and the target categories,
the proposed \proposed framework adopts the two-step approach that takes advantage of both the textual similarity encoded in a text embedding space and the discriminative power of a neural text classifier.
Our empirical evaluation demonstrates that \proposed successfully identifies out-of-category documents in various target scenarios.
In conclusion, \proposed can be practically used in many real-world applications for filtering out the documents that are less relevant to inlier categories or user-interested topics, only requiring the minimum guidance.

\smallsection{Acknowledgement}
This work was supported by the NRF grant (No. 2020R1A2B5B03097210), the IITP grant (No. 2018-0-00584, 2019-0-01906), US DARPA KAIROS Program (No. FA8750-19-2-1004), SocialSim Program (No.  W911NF-17-C-0099), INCAS Program (No. HR001121C0165), National Science Foundation (IIS-19-56151, IIS-17-41317, IIS 17-04532), and the Molecule Maker Lab Institute: An AI Research Institutes program (No. 2019897).

\bibliographystyle{IEEEtran}
\bibliography{BIB/bibliography}

\begin{thebibliography}{10}
\providecommand{\url}[1]{#1}
\csname url@samestyle\endcsname
\providecommand{\newblock}{\relax}
\providecommand{\bibinfo}[2]{#2}
\providecommand{\BIBentrySTDinterwordspacing}{\spaceskip=0pt\relax}
\providecommand{\BIBentryALTinterwordstretchfactor}{4}
\providecommand{\BIBentryALTinterwordspacing}{\spaceskip=\fontdimen2\font plus
\BIBentryALTinterwordstretchfactor\fontdimen3\font minus
  \fontdimen4\font\relax}
\providecommand{\BIBforeignlanguage}[2]{{%
\expandafter\ifx\csname l@#1\endcsname\relax
\typeout{** WARNING: IEEEtran.bst: No hyphenation pattern has been}%
\typeout{** loaded for the language `#1'. Using the pattern for}%
\typeout{** the default language instead.}%
\else
\language=\csname l@#1\endcsname
\fi
#2}}
\providecommand{\BIBdecl}{\relax}
\BIBdecl

\bibitem{hauskrecht2013outlier}
M.~Hauskrecht, I.~Batal, M.~Valko, S.~Visweswaran, G.~F. Cooper, and
  G.~Clermont, ``Outlier detection for patient monitoring and alerting,''
  \emph{JBI}, vol.~46, no.~1, pp. 47--55, 2013.

\bibitem{kannan2017outlier}
R.~Kannan, H.~Woo, C.~C. Aggarwal, and H.~Park, ``Outlier detection for text
  data,'' in \emph{SDM}, 2017, pp. 489--497.

\bibitem{yin2016model}
J.~Yin and J.~Wang, ``A model-based approach for text clustering with outlier
  detection,'' in \emph{ICDE}, 2016, pp. 625--636.

\bibitem{zhuang2017identifying}
H.~Zhuang, C.~Wang, F.~Tao, L.~Kaplan, and J.~Han, ``Identifying semantically
  deviating outlier documents,'' in \emph{EMNLP}, 2017, pp. 2748--2757.

\bibitem{fouche2020mining}
E.~Fouch{\'e}, Y.~Meng, F.~Guo, H.~Zhuang, K.~B{\"o}hm, and J.~Han, ``Mining
  text outliers in document directories,'' in \emph{ICDM}, 2020, pp. 152--161.

\bibitem{mohotti2020efficient}
W.~A. Mohotti and R.~Nayak, ``Efficient outlier detection in text corpus using
  rare frequency and ranking,'' \emph{TKDD}, vol.~14, no.~6, pp. 1--30, 2020.

\bibitem{breunig2000lof}
M.~M. Breunig, H.-P. Kriegel, R.~T. Ng, and J.~Sander, ``Lof: identifying
  density-based local outliers,'' in \emph{SIGMOD}, 2000, pp. 93--104.

\bibitem{sathe2016subspace}
S.~Sathe and C.~C. Aggarwal, ``Subspace outlier detection in linear time with
  randomized hashing,'' in \emph{ICDM}, 2016, pp. 459--468.

\bibitem{ruff2019self}
L.~Ruff, Y.~Zemlyanskiy, R.~Vandermeulen, T.~Schnake, and M.~Kloft,
  ``Self-attentive, multi-context one-class classification for unsupervised
  anomaly detection on text,'' in \emph{ACL}, 2019, pp. 4061--4071.

\bibitem{hendrycks2020pretrained}
D.~Hendrycks, X.~Liu, E.~Wallace, A.~Dziedzic, R.~Krishnan, and D.~Song,
  ``Pretrained transformers improve out-of-distribution robustness,'' in
  \emph{ACL}, 2020, pp. 2744--2751.

\bibitem{lee2020multi}
D.~Lee, S.~Yu, and H.~Yu, ``Multi-class data description for
  out-of-distribution detection,'' in \emph{KDD}, 2020, pp. 1362--1370.

\bibitem{mikolov2013distributed}
T.~Mikolov, I.~Sutskever, K.~Chen, G.~Corrado, and J.~Dean, ``Distributed
  representations of words and phrases and their compositionality,'' in
  \emph{NeurIPS}, 2013.

\bibitem{meng2019spherical}
Y.~Meng, J.~Huang, G.~Wang, C.~Zhang, H.~Zhuang, L.~Kaplan, and J.~Han,
  ``Spherical text embedding,'' \emph{NeurIPS}, vol.~32, 2019.

\bibitem{manolache2021date}
A.~Manolache, F.~Brad, and E.~Burceanu, ``Date: Detecting anomalies in text via
  self-supervision of transformers,'' in \emph{NAACL-HLT}, 2021.

\bibitem{moon2021masker}
S.~J. Moon, S.~Mo, K.~Lee, J.~Lee, and J.~Shin, ``Masker: Masked keyword
  regularization for reliable text classification,'' in \emph{AAAI}, 2021.

\bibitem{meng2018weakly}
Y.~Meng, J.~Shen, C.~Zhang, and J.~Han, ``Weakly-supervised neural text
  classification,'' in \emph{CIKM}, 2018, pp. 983--992.

\bibitem{meng2020text}
Y.~Meng, Y.~Zhang, J.~Huang, C.~Xiong, H.~Ji, C.~Zhang, and J.~Han, ``Text
  classification using label names only: A language model self-training
  approach,'' in \emph{EMNLP}, 2020, pp. 9006--9017.

\bibitem{huang2020weakly}
J.~Huang, Y.~Meng, F.~Guo, H.~Ji, and J.~Han, ``Weakly-supervised aspect-based
  sentiment analysis via joint aspect-sentiment topic embedding,'' in
  \emph{EMNLP}, 2020, pp. 6989--6999.

\bibitem{meng2020hierarchical}
Y.~Meng, Y.~Zhang, J.~Huang, Y.~Zhang, C.~Zhang, and J.~Han, ``Hierarchical
  topic mining via joint spherical tree and text embedding,'' in \emph{KDD},
  2020, pp. 1908--1917.

\bibitem{vilnis2015word}
L.~Vilnis and A.~McCallum, ``Word representations via gaussian embedding,'' in
  \emph{ICLR}, 2015.

\bibitem{vendrov2016order}
I.~Vendrov, R.~Kiros, S.~Fidler, and R.~Urtasun, ``Order-embeddings of images
  and language,'' in \emph{ICLR}, 2016.

\bibitem{ganea2018hyperbolic}
O.~Ganea, G.~B{\'e}cigneul, and T.~Hofmann, ``Hyperbolic entailment cones for
  learning hierarchical embeddings,'' in \emph{ICML}, 2018, pp. 1646--1655.

\bibitem{pennington2014glove}
J.~Pennington, R.~Socher, and C.~D. Manning, ``Glove: Global vectors for word
  representation,'' in \emph{EMNLP}, 2014, pp. 1532--1543.

\bibitem{hinton2015distilling}
G.~Hinton, O.~Vinyals, and J.~Dean, ``Distilling the knowledge in a neural
  network,'' \emph{arXiv preprint arXiv:1503.02531}, 2015.

\bibitem{xie2016unsupervised}
J.~Xie, R.~Girshick, and A.~Farhadi, ``Unsupervised deep embedding for
  clustering analysis,'' in \emph{ICML}, 2016, pp. 478--487.

\bibitem{wang2019effective}
S.~Wang, Y.~Zeng, X.~Liu, E.~Zhu, J.~Yin, C.~Xu, and M.~Kloft, ``Effective
  end-to-end unsupervised outlier detection via inlier priority of
  discriminative network.'' in \emph{NeurIPS}, 2019, pp. 5960--5973.

\bibitem{shang2018automated}
J.~Shang, J.~Liu, M.~Jiang, X.~Ren, C.~R. Voss, and J.~Han, ``Automated phrase
  mining from massive text corpora,'' \emph{TKDE}, vol.~30, no.~10, pp.
  1825--1837, 2018.

\bibitem{sennrich2016neural}
R.~Sennrich, B.~Haddow, and A.~Birch, ``Neural machine translation of rare
  words with subword units,'' in \emph{ACL}, 2016, pp. 1715--1725.

\bibitem{kudo2018sentencepiece}
T.~Kudo and J.~Richardson, ``Sentencepiece: A simple and language independent
  subword tokenizer and detokenizer for neural text processing,'' in
  \emph{EMNLP Demo}, 2018, pp. 66--71.

\bibitem{kim-2014-convolutional}
Y.~Kim, ``Convolutional neural networks for sentence classification,'' in
  \emph{EMNLP}, Oct. 2014, pp. 1746--1751.

\end{thebibliography}

\end{document}